\documentclass[apj]{emulateapj}
\usepackage{graphicx}
\usepackage[dvips]{color}

\usepackage{graphicx,amsmath,amssymb,epsfig}

\usepackage{natbib}
\newcommand{\be}{\begin{equation}}
\newcommand{\ee}{\end{equation}}
\newcommand{\ba}{\begin{eqnarray}}
\newcommand{\ea}{\end{eqnarray}}

\renewcommand{\vec}[1]{\boldsymbol{#1}}

\def\d{{\rm d}}

\def\lsim{\raise0.3ex\hbox{$\;<$\kern-0.75em\raise-1.1ex\hbox{$\sim\;$}}}
\def\gsim{\raise0.3ex\hbox{$\;>$\kern-0.75em\raise-1.1ex\hbox{$\sim\;$}}}

\def\theta{\vartheta}

\renewcommand{\vec}[1]{\boldsymbol{#1}}

\shortauthors{Savchenko et al.}
\shorttitle{Imprint of a 2\,Myr old source on the cosmic ray anisotropy}

\begin{document}
\author{V.~Savchenko$^{1}$, M.~Kachelrie\ss$^{2}$, D.~V.~Semikoz$^{1}$}
\affil{
$^{1}$AstroParticle and Cosmology (APC), Paris, France\\
$^{2}$Institutt for fysikk, NTNU, Trondheim, Norway}

\title{Imprint of a 2\,Million year old source on the cosmic ray anisotropy}

\begin{abstract}
We study numerically the anisotropy of the cosmic ray (CR) flux emitted by a 
single source calculating the trajectories of individual CRs. We show that 
the contribution of a single source to the observed  anisotropy is determined 
solely by the fraction the source  contributes  to the total CR intensity, 
its age and its distance, and does not depend on the CR energy at late times. 
Therefore the observation of a constant dipole
anisotropy indicates that a single source dominates the CR flux in 
the corresponding energy range.  A natural explanation for the plateau
between 2--20\,TeV observed in the CR anisotropy is thus the presence 
of a single, nearby source.
For the source age of 2\,Myr, as suggested by the explanation of the 
antiproton and positron data from PAMELA and AMS-02 through a local source,
we determine the source distance as $\sim 200$\,pc.
Combined with the contribution of the global CR sea calculated in the
escape model, we can explain qualitatively the data for the dipole  anisotropy.
Our results suggest that the assumption of a smooth CR source distribution
should be abandoned between $\simeq 200$\,GeV and 1\,PeV. 
\end{abstract}


\maketitle

%%%%%%%%%%%%%%%%%%%%%%%%%%%%%%%%%%%%%%%%%%%%%%%%%%%%%%%%%%%%%%%%%%%%%%
\section{Introduction}
\label{Introduction}
%%%%%%%%%%%%%%%%%%%%%%%%%%%%%%%%%%%%%%%%%%%%%%%%%%%%%%%%%%%%%%%%%%%%%%

The observed intensity of cosmic rays (CR) is characterised by a 
large degree of isotropy up to the highest energies. 
The measured dipole anisotropy $\delta$ increases up $E\sim 2$\,TeV,
forms an approximately energy independent plateau between $E\sim 2$--20\,TeV,
before the amplitude decreases again. Finally, at energies 
$E\gsim 100$\,TeV, the anisotropy grows fast with energy.
Such a behavior of the dipole anisotropy $\delta$ as function of energy
seems at first sight difficult to reconcile with  diffusive CR propagation: 
In this picture, the scattering of CRs on 
inhomogeneities of the Galactic magnetic field (GMF) converts
their trajectories to a random walk, erasing thereby most directional 
information. Cosmic rays are most effectively scattered by those 
turbulent field modes whose wave-length equal their Larmor
radii. Therefore the scattering rate is energy dependent and determined by 
the fluctuation spectrum of the turbulent GMF. 
As result, both the diffusion coefficient and the CR anisotropy are expected 
to increase with energy. This picture is supported e.g.\ by recent data from 
the AMS-02 experiment for the B/C ratio which are consistent with the simple
power-law $D(E)\propto E^{0.31}$ up to the rigidity 1.8\,TV.

In the diffusion approximation,  Fick's law is valid and the net CR current
$\vec j(E)$ is determined by the gradient of the CR number density 
$n(E)=\d N/(\d E \d V)$ and the diffusion tensor $D_{ab}(E)$ as
$j_a = -D_{ab}\nabla_b n$. The dipole vector $\vec \delta$ of the CR
intensity $I= c/(4\pi)n$ follows then as
\begin{equation} \label{delta_diff}
 \delta_a %\equiv \frac{I_{\max}-I_{\min}}{I_{\max}+I_{\min}}
          =  \frac{3}{c} \frac{j_a}{n} 
          = - \frac{3D_{ab}}{c}\frac{\nabla_b n}{n} \,.
\end{equation}
Within this approximation, the diffusion tensor $D_{ab}$ is an external input. 
The traditional approach to determine $D_{ab}$  as a function of the assumed 
magnetic field uses kinetic theory, see e.g.~\citet{bookBerezinskii}.
Such an analytical approach allows one to connect both $D_{ab}$ and 
$\delta_a$ to the spectrum of magnetic field fluctuations, involves however 
the use of approximations.

An alternative method uses the trajectories of individual CRs calculated 
numerically solving the Lorentz equation in turbulent and regular magnetic 
fields. This approach is  numerically expensive and has been restricted 
mainly to the calculation of the  diffusion tensor 
$D_{ab}$~\citep{1999ApJ...520..204G,Casse:2001be,Giacinti:2012ar}.
Since the anisotropy in the astrophysically interesting cases is small, 
its calculation requires a much larger number of CR trajectories. 
Therefore the determination of the CR anisotropy expected from a smooth
source distribution has not been feasible using this method directly. 
Instead, the approximation proposed by \citet{Karakula:1972na}
has been widely used in the high-energy range, $E\gsim 10^{18}$\,eV. 
In this  letter, we concentrate on the case of a single source and show 
that the calculation of the anisotropy from first principles is possible
for an interesting range of parameters.

The  propagation of CRs can be divided into three different regimes: 
The free streaming, the anisotropic and the quasi-Gaussian diffusion of CRs.
In the first regime, the change of the particle momentum is 
small. The intermediate regime of  anisotropic diffusion was explored first 
by \citet{Giacinti:2012ar,Giacinti:2013wbp}; a study of the anisotropy in 
this regime will be presented elsewhere. % in~\citep{long}.
 Here, we concentrate on the case of  quasi-Gaussian diffusion and show that
then the dipole anisotropy of a single source with age $T$ and distance
$R$ is given by $\delta=3R/(2cT)$, independent of the regular and turbulent
magnetic field. As an application of this result, we consider the observed
CR dipole anisotropy. We propose as explanation for the plateau in the
observed anisotropy between 2--20\,TeV that a single nearby source dominates 
the CR flux in  this energy range. At lower energies, the anisotropy 
decreases because the fraction of CRs contributed by the single source 
goes down.
We show that the anisotropy data between 200\,GeV and 4\,PeV can be 
naturally explained by a local CR source with the age of 2\,Myr
and distance $\sim 200$\,pc, consistent with the characteristics of the single 
source determined in~\citet{LS} for an
explanation of the antiproton and positron data from PAMELA and AMS-02.

\begin{figure*}
\includegraphics[width=0.75\columnwidth,angle=-90,origin=c]{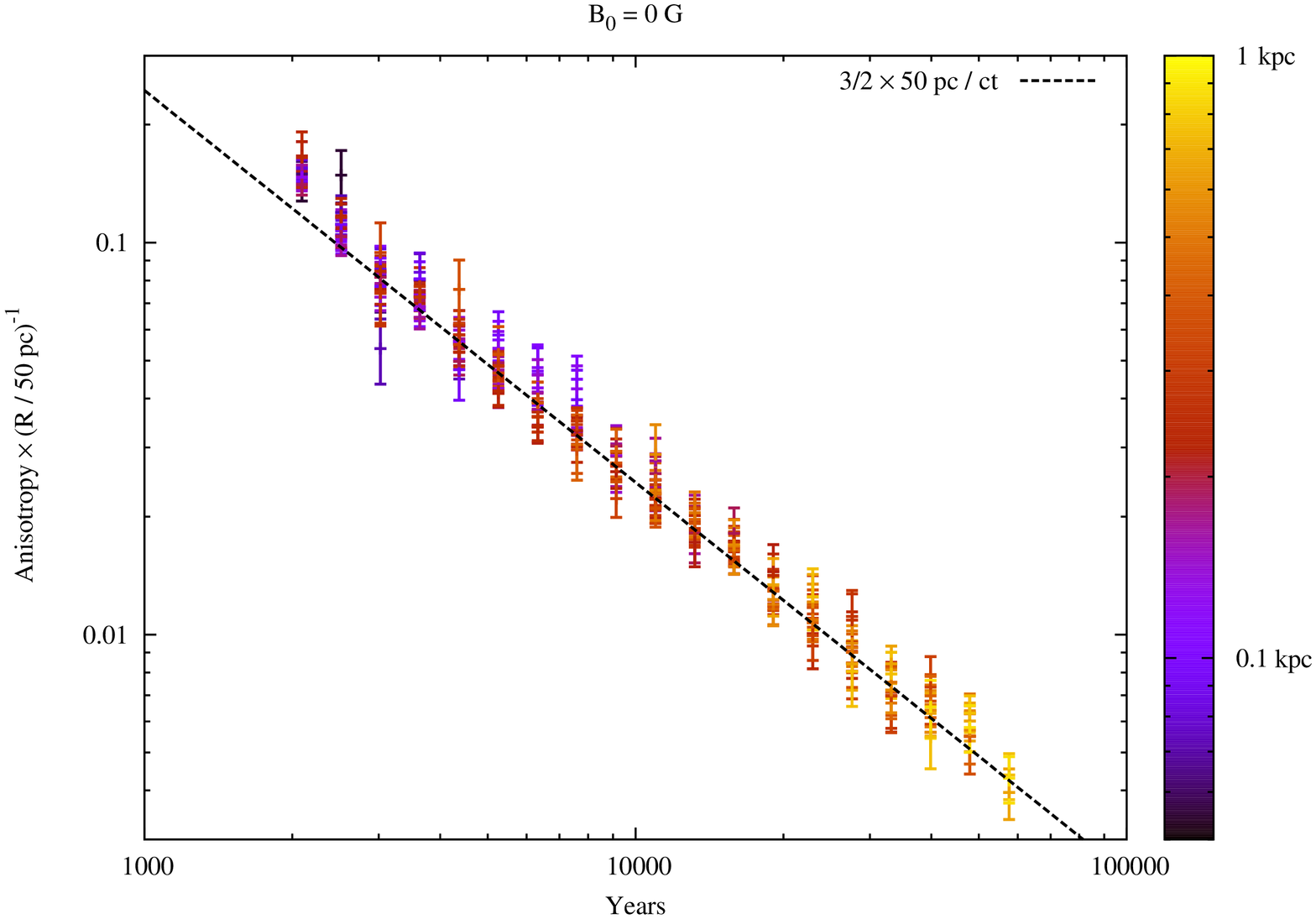}
\hskip0.2cm
\includegraphics[width=0.75\columnwidth,angle=-90,origin=c]{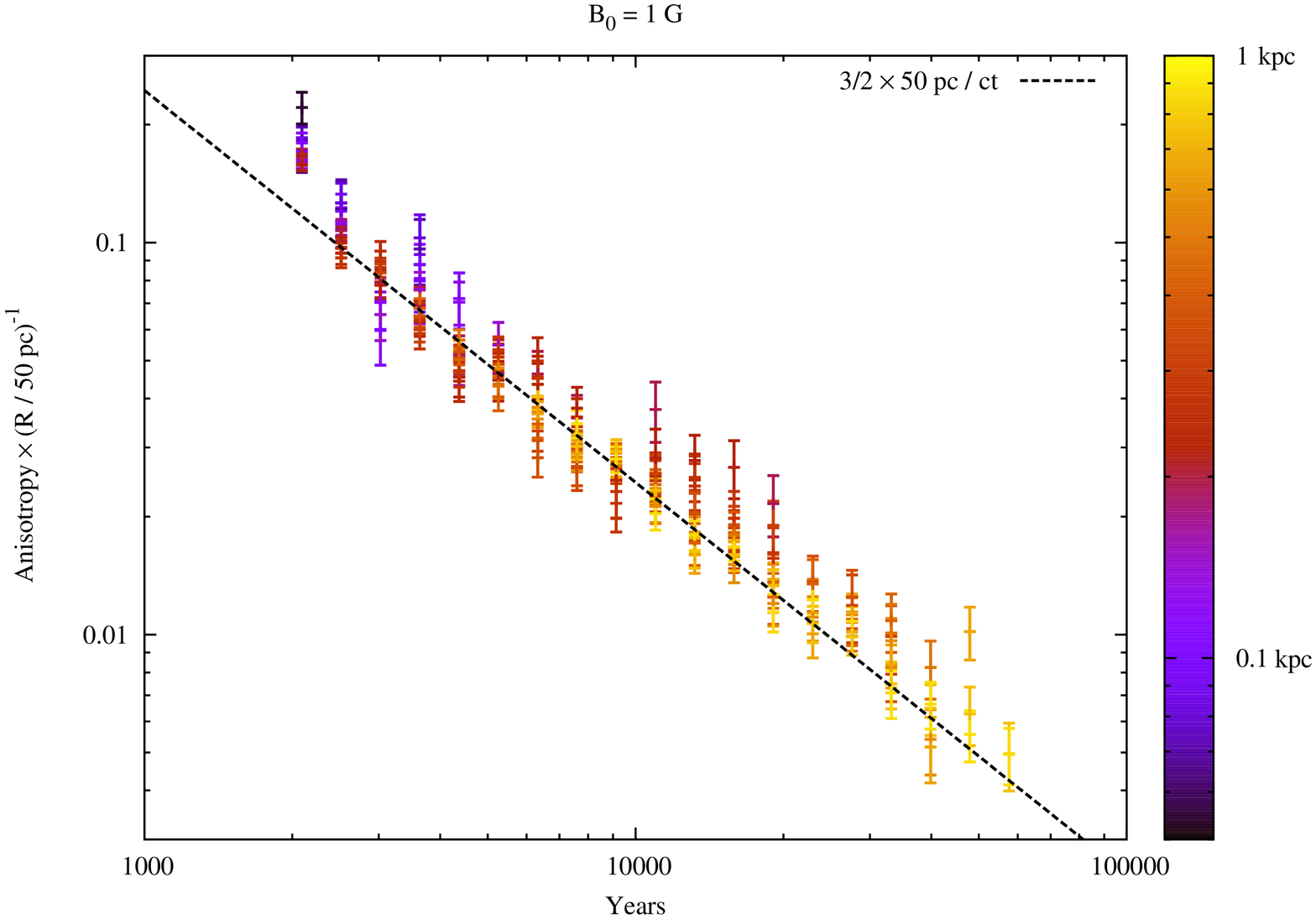}
\vskip-1.25cm
\caption{The rescaled flux anisotropy $A$ of single sources at various 
distances as function of time. The distance is indicated by the color code 
in kpc.  Left panel: 
  turbulent magnetic field $B_{\rm rms}=0.3\,\mu$G; right panel: 
$B_0=1\,\mu$G and $B_{\rm rms}=0.3\,\mu$G.
\label{Asource}}
\end{figure*}

%%%%%%%%%%%%%%%%%%%%%%%%%%%%%%%%%%%%%%%%%%%%%%%%%%%%%%%%%%%%%%%%%%%%%%
\section{Anisotropy in the quasi-Gaussian regime}
\label{gaussian}

We avoid the limitations of the diffusion approximation by 
calculating the trajectories of individual CRs in a given regular
and turbulent magnetic field using the numerical code described 
by \citet{Giacinti:2011uj,Giacinti:2011ww}. 
We use as spectrum $\mathcal{P}(\vec k)$ 
of the magnetic field fluctuations  isotropic Kolmogorov turbulence,
$\mathcal{P}(k)\propto k^{-\alpha}$ with $\alpha=5/3$ and set the
maximum length $l_{\rm max}$ equal to 25\,pc. For the numerical calculations, 
we use nested grids. This allows us to choose an effective $l'_{\min}$ 
sufficiently small compared to the Larmor radius %$R_{\rm L}=cp/(eB)$ 
considered. 

We model the CR source as an instantaneous injection of CRs at a single
point. Since we are interested in the generic behavior of CR diffusion on
relatively small length scales, we use not a concrete GMF model but approximate
its regular component by a uniform field. We measure the momentum distribution
$f(\vec p)$ of CRs crossing a set of spheres with radii between 1\,pc and 
1\,kpc centered on the CR source.  
We verify that in the quasi-Gaussian regime the distribution is compatible with 
a dipole, and compute the anisotropy $A$ of the flux of CRs crossing each 
sphere\footnote{Note that the anisotropy $A\equiv (F_{\max}- F_{\min})/(F_{\max}+F_{\min})$ of the CR 
  flux $F$ is connected to the anisotropy $\delta= (I_{\max}-I_{\min})/(I_{\max}+I_{\min})$ of the CR intensity $I$ by  $\delta=3A/2$.}.
The statistical error of the anisotropy estimate is a factor  $1/A^2$
larger than the error of the corresponding flux estimate. Thus even for 
relatively large flux anisotropies,  $A \simeq 0.01$, the number of 
trajectories 
and hence the CPU time required is increased by a factor $10^{4}$. The 
average total computing time used for a single run was about 10 CPU years.
To further reduce the statistical uncertainty, we have averaged the
anisotropy over regions selected according to the symmetry of the 
regular magnetic field $B_0$: spherical symmetry for $B_0=0$ and 
axial symmetry for $B_0={\rm const}$.

In the left panel of Fig.~\ref{Asource}, we show the flux anisotropy $A$ 
of single sources at various distances as function of time
for a purely turbulent magnetic field with strength $B_{\rm rms}=0.3\,\mu$G. 
We choose the times and length scales sufficiently large such that the 
CR propagation proceeds in the quasi-Gaussian regime,  $t\gg 2D/c^2$
and  $l\gg l_{\rm coh}$. For a purely turbulent field, 
CRs should perform a random walk. Thus their density
is Gaussian, and as a result, Eq.~(\ref{delta_diff}) simplifies \citep{1971ApL.....9..169S} to
\be \label{single}
 A = \frac{R}{cT} \simeq 3.3 \times 10^{-4} \left(\frac{R}{200\,{\rm pc}}\right)
                                 \left(\frac{T}{2\,{\rm Myr}}\right)^{-1}
\,.
\ee
The flux anisotropy $A$ shown in  Fig.~\ref{Asource} for sources at varying
distances is rescaled to the value of  the nearest source at 50\,pc
assuming the validity of Eq.~(\ref{single}). Within the numerical
precision of our results, the obtained values for $A$ follow nicely
the analytical result.

Next we add a regular field to the turbulent field. In this case,
CRs diffuse faster parallel than perpendicular to the regular field lines 
and  one may wonder, if the simple formula~(\ref{single}) for the dipole
anisotropy still holds. Restricting ourselves again to the case
$t\gg 2D/c^2$ and  $l\gg l_{\rm coh}$, we confirm numerically that CR propagate 
also with a non-zero regular field in the quasi-Gaussian regime. 
Now the widths $4D_it$ of 
the three-dimensional  Gaussian describing the CR density 
$n(E,\vec x)$ are given by the corresponding components of the diffusion 
tensor. Consequently, we expect that the regular field leads to strong 
deviations from a radially symmetric number density of CRs but does not 
influence the anisotropy.
In the right panel of Fig.~\ref{Asource}, we show our results
for the flux anisotropy $A$ in the presence of both a turbulent and
regular field, choosing as an example the field-strengths 
$B_0=1\,\mu$G and $B_{\rm rms}=0.3\,\mu$G. Comparing the results
in the left (without) and the right (with regular field) panel,
it is clear that the regular field has no impact on the anisotropy.

\begin{figure}
\includegraphics[width=0.7\columnwidth,angle=270]{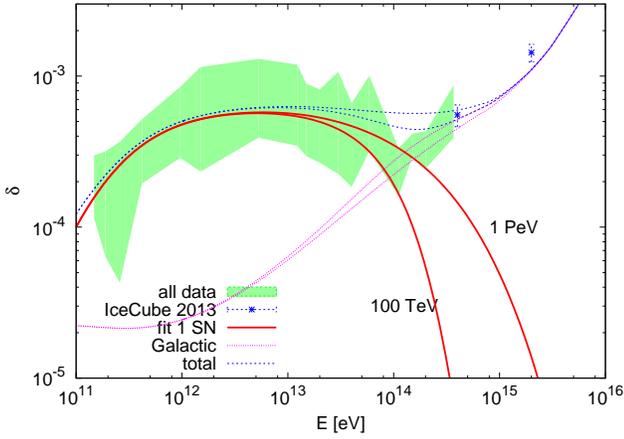}
\caption{Lower and upper limit (green band) %from the experimental data 
on the dipole  anisotropy and data from IceCube (blue errorbars) compared 
to the contribution from the local  source (red, for two values of $E_{\max}$) 
and from the average CR sea  (magenta)  as function of energy.
\label{Adata}}
\end{figure}

\begin{figure}
\includegraphics[width=0.7\columnwidth,angle=270]{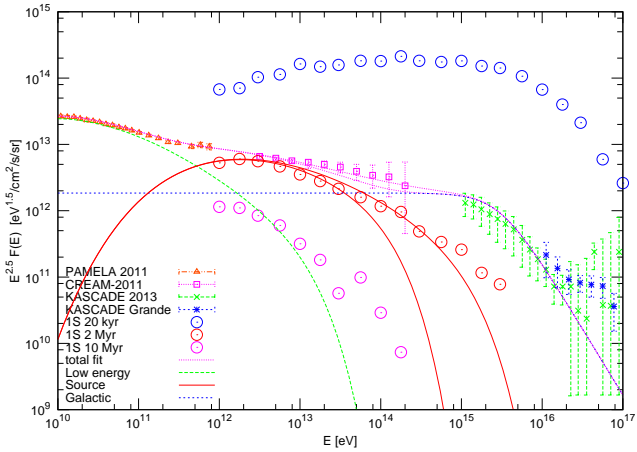}
\caption{Contribution from average Galactic (blue line), from the local 
source (red lines, for cutoff $E_{\max}=10^{14}$ and $10^{15}$\,eV), and old
local sources (green line)
to the CR proton intensity together with experimental data (errorbars).
Additionally, the flux from the local source is shown by dots for three
source ages.
\label{flux}}
\end{figure}

%%%%%%%%%%%%%%%%%%%%%%%%%%%%%%%%%%%%%%%%%%%%%%%%%%%%%%%%%%%%%%%%%%%%%%
\section{Interpretation of the observed CR  anisotropy}
\label{data}

In Fig.~\ref{Adata}, we show as band the range of experimental data 
collected by~\citet{DiSciascio:2014jwa} on the magnitude of the CR dipole 
anisotropy%\footnote{The experimental values $\tilde\delta$ are connected to the actual dipole amplitude $\delta$ by $\tilde\delta= \delta_\perp/\langle\cos(\rm dec)\rangle$, where $\delta_\perp$ is the dipole component in the equatorial plane and $\langle\cos(\rm dec)\rangle$ the average declination of the events used in the harmonic analysis based only on R.A.} 
$\delta$ together with data 
from %EAS-TOP (orange errorbars) \citep{Aglietta:2009mu} and 
IceCube (blue errorbars) \citep{Abbasi:2011zka}. 
The anisotropy grows as function of energy until $E\sim 2$\,TeV,
remains approximately constant in the range 2--20\,TeV, before
it decreases again. Starting from 100\,TeV the anisotropy increases,
with a rate which is consistent with one determined in the escape
model of \citet{escape1,escape2}.

In the standard diffusion picture, the anisotropy is connected
to the gradient of the global CR sea density and the diffusion 
coefficient. Both quantities increase with energy and, consequently,
the energy dependence shown in Fig.~\ref{Adata} is difficult to
explain. Our results from the previous section suggest to connect
the plateau in the dipole anisotropy between 2--20\,TeV to the presence
of a single nearby source. 
We determine the relative contribution to the total CR intensity from
the nearby source, $f_s= I_s(E)/I_{\rm tot}$, and from the global CR sea 
using the fluxes from \citet{LS}. The contribution of the local source to 
the dipole anisotropy follows then as $\delta_s= 3f_i R/(2cT)$.
This anisotropy is shown with red lines in Fig.~\ref{Adata}.
%Moreover, these results fixed the age of the source quite precisely,
%$T\simeq 2$\,Myr, while its distance $R$ remained rather uncertain.  
%Inverting the anisotropy formula, we find 
%$R\simeq 200\,{\rm pc}\times \delta/(5\times 10^{-4})$.
Combined with $T\simeq 2$\,Myr as age of the source,  we can determine
its distance as $R\simeq 200\,{\rm pc}\times \delta_s/(5\times 10^{-4})$.

Since the spread of the experimental data on the dipole is considerable, let 
us consider for illustration the following two extreme situations: First, 
the true dipole may be close to the lower limit of the measured range, 
$\delta\sim 3\times 10^{-4}$. In this case, the plateau extends up
to %the first measurement of EAS-TOP at $1.1\times 10^{14}$\,eV, the
$\simeq 10^{14}$\,eV, the
maximal energy of CRs accelerated in the local source is unrestricted
and the source distance could be as low as 100\,pc. 
The following rise of the dipole is then explained by the contribution of the 
global CR sea which we calculate in the escape model of \citet{escape1,escape2}.
In the other extreme, the true dipole may be close to the upper limit of 
the measurements, $\delta\sim 10^{-3}$. In this case, we have to introduce a
cutoff $\exp(-E/E_{\max})$ in the energy of CRs  accelerated in the local 
source such that their contribution to the total CR intensity and thus to 
the observed dipole decreases towards the EAS-Top value.

We can constrain these extreme choices considering also the total CR proton 
flux.  In Fig.~\ref{flux}, we compare experimental results for the 
proton flux from CREAM~\citep{Yoon:2011aa}, 
PAMELA~\citep{Adriani:2011cu}, KASCADE and KASCADE-Grande~\citep{Apel:2011mi},
to the proton flux from the local source shown by dots and the average
global CR sea at low energies (green line).
Larger maximal values of $\delta$ require a 
lower value of $E_{\max}$, making the source spectrum more ``bumpy''.  
Such a spectral distortion in the total proton flux is avoided for
values of $E_{\max}\gsim 10^{14}$\,eV,  what also ensures that the antiproton 
data from AMS-02 can be explained.

Let us now comment on the decrease of the measured dipole anisotropy 
at low-energies, $\lsim 2$\,TeV. A natural explanation for this decrease
is a small off-set of the Earth with respect to magnetic field line going 
through the source. The perpendicular extension of the elongated volume 
filled with CRs by the source decreases with energy. 
Thus the lower end of the plateau is determined by the perpendicular
distance of the Earth to the magnetic field line going through the
source and is therefore a free parameter.
In Fig.~\ref{flux}, we show the proton flux from the local source as
a red line using $E_{\min}= 10^{12}$\,eV and $E_{\max}= 10^{14}$\,eV and 
$10^{15}$\,eV  as low- and  high-energy cutoffs, respectively.

The difference at low energies between the observed CR flux and the one 
from the local source  should be contributed by old, local sources. This
contribution is
shown for illustration in Fig. 3 as a green line, obtained by subtracting
the two fluxes. %\citet{2015ApJ...800...71F} 
Fry et~al.\ (2015) showed that their age should be
$T>14$\,Myr. Therefore the high-energy CRs emitted by  these sources escaped
already, what might explain the suppression of their combined CR fluxes
above $10^{11}$\,eV. Their individual dipoles $\delta_i\propto f_i/T$ are 
smaller, should partially cancel in their sum, and we therefore neglect
their contribution to the total dipole amplitude in Fig.~\ref{Adata}.
%
%The average global CR sea at low energies (green line) consists of 
%several old sources, $T\gg 2$\,Myr. Their individual dipoles are therefore
%smaller, should partially cancel in their sum, and we therefore neglect
%their contribution to the total dipole amplitude in Fig.~\ref{Adata}.
%The high-energy CRs emitted by these sources 
%escaped already, explaining the suppression of their combined
%CR fluxes above $10^{11}$\,eV. 
Since the number $N$ of sources contributing
is large, the approximation of a smooth source distribution and thus also
the standard approach used e.g.\ by 
\citet{Strong:2007nh,Evoli:2008dv,2011A&A...526A.101P} may be justified
in this energy range. The transition between the single and many source
regimes around $E\sim 2\times 10^{11}$\,eV  may be connected to the spectral
breaks observed by PAMELA.

The phase of the first harmonics is almost constant up to 100\,TeV 
energy~\citep{DiSciascio:2014jwa}, and changes thereafter 
fast. Such a behavior is natural in our model: At all energies 
$E\lsim 100$\,TeV, the anisotropy is dominated 
by local source(s) which are located  preferentially along the local 
magnetic field lines. As result, the dipole phase expected %in our model  
does not change its  value up to 100\,TeV, where the Galactic sea of CRs 
starts to dominate and thus the dipole  direction is given by 
Eq.~(\ref{delta_diff}).

The presence of a nearby source dominating the CR flux and dipole in the
TeV range is natural to expect~\citep{LS}:
The average rate of SN explosions in the Milky Way 
%disk volume $V_{\rm disk}\simeq 100$--300\,kpc$^3$ is ${\cal R}_{SN} \simeq (1$--$3)\times 10^{-2}$\,yr$^{-1}$, or 
is one SN per (0.3--3)$\times 10^4$\,yr per kpc$^3$. Cosmic rays in the 
TeV energy range fill a 100\,pc wide, kpc long volume directed 
along the regular GMF lines for a time of a Myr. It is thus likely   
that one SN has exploded during the last Myr within this volume.
%elongated along the GMF lines passing near the Solar system. 

Such a source is consistent with the explanation for the deposition of 
$^{60}$Fe isotopes in the %million years old 
deep ocean crust by the passage of an expanding shell of a 2\,Myr old 
supernova remnant through the Solar System~\citep{Knie:1999zz,Benitez:2002jt, Fry:2014yqa}. The distance determined by ~\citet{Fry:2014yqa} e.g.\ for the
case of a $25M_\odot$ core-collapse supernova is $\simeq 130$\,pc. Relaxing 
some of the assumptions like an isotropic emission, this estimate may be 
compatible with our value $R\sim 200$\,pc in case of a low dipole amplitude. 
Moreover, a single local source dominating the local CR proton
spectrum explains the known differences in the slopes between  CR protons 
and nuclei and the anomalous energy dependence of  the spectra of positrons 
and  antiprotons~\citep{LS}.
% as shown in \citet{LS}.

%%%%%%%%%%%%%%%%%%%%%%%%%%%%%%%%%%%%%%%%%%%%%%%%%%%%%%%%%%%%%%%%%%%%%%
\section{Discussion and Conclusions}
\label{conclusions}
%%%%%%%%%%%%%%%%%%%%%%%%%%%%%%%%%%%%%%%%%%%%%%%%%%%%%%%%%%%%%%%%%%%%%%

The standard approach to CR propagation using either analytical~\citep{bookBerezinskii} or numerical methods~\citep{galprop,Strong:2007nh,Evoli:2008dv} assumes a CR source distribution that is smooth in time and space. 
\citet{Blasi:2011fi,Blasi:2011fm} based on earlier work by \citet{1979ApJ...229..424L} studied the fluctuations in the CR density and the dipole induced by the stochastic nature of CR sources. Their results for the fluctuations  diverge in the limit of small source distances and ages, $R_{\min},T_{\min}\to 0$.
These divergences indicate that a statistical description of quantities like the CR dipole anisotropy in terms of its ensemble average and variance is not adequate, since the results depend strongly on the actual  properties of the nearest 
source(s) in a given realization.

The strongly anisotropic diffusion of CRs that is typical for the GMF parameters favored by the escape model enhances these fluctuations: The majority of recent nearby sources is not connected to us by the regular field, and their contribution to the local CR intensity is therefore suppressed. In contrast, the flux of the single (or the few) sources active in the last few million years that are located in the volume aligned with the GMF lines passing near the Solar system is strongly enhanced. These effects prevent the mixing of CRs with energies 1-100 TeV of various sources into a ``global, average CR sea,'' and undermine thereby the basic assumption of a smooth, global CR distribution inherent in most approaches to Galactic CR physics.

We have shown that the dipole anisotropy in the CR flux emitted by a single 
source is independent of the  turbulent and regular magnetic field and of 
the CR energy in the quasi-Gaussian regime. %, i.e.\ for  $t\gg 2D/c^2$
%and  $l\gg l_{\rm coh}$. 
In particular, we have shown that the simple 
formula $A =R/(cT)$ for the dipole anisotropy holds also including a 
regular field.
As an application we have considered the experimental data on the dipole
anisotropy. We have argued that the approximately energy-independent
plateau in the anisotropy around 2-20\,TeV can be explained by the
presence of a nearby CR source. The  age and the distance of this source
are compatible with the one required to explain the ``anomalies'' in the 
CR intensities of protons, antiprotons and positrons. If the source is
also responsible for the observed $^{60}$Fe overabundance in the million 
year old ocean crust, its distance should 100--200\,pc, favoring thus
values of the dipole anisotropy close to the lower end of the measured
values.

Finally we note that our results can be used also to study
the dipole anisotropy of the electron flux, constraining e.g.\
the contribution from recent pulsars, if energy losses are 
taken into account.

%%%%%%%%%%%%%%%%%%%%%%%%%%%%%%%%%%%%%%%%%%%%%%%%%%%%%%%%%%%%%%%%%%%%%%
\acknowledgments

We would like to thank Evgeny Babichev for useful discussions. 
%We acknowledge the support of France Grilles for providing computing 
We acknowledge France Grilles for providing computing 
resources on the French National Grid Infrastructure and  the 
Fran\c{c}ois Arago Centre at APC. % for local computing resources.
MK is grateful to the Theory Group at APC for hospitality.  
The work of DS was supported in part by the grant RFBR 
\# 13-02-12175-ofi-m.

%%%%%%%%%%%%%%%%%%%%%%%%%%%%%%%%%%%%%%%%%%%%%%%%%%%%%%%%%%%%%%%%%%%%%%

%\bibliography{CRprop,CRprop2}

\begin{thebibliography}{26}
\expandafter\ifx\csname natexlab\endcsname\relax\def\natexlab#1{#1}\fi

\bibitem[{Abbasi {et~al.}(2012)}]{Abbasi:2011zka}
Abbasi, R., {et~al.} 2012, Astrophys.J., 746, 33, 1109.1017

\bibitem[{Adriani {et~al.}(2011)}]{Adriani:2011cu}
Adriani, O., {et~al.} 2011, Science, 332, 69, 1103.4055

\bibitem[{Apel {et~al.}(2011)}]{Apel:2011mi}
Apel, W., {et~al.} 2011, Phys.Rev.Lett., 107, 171104, 1107.5885

\bibitem[{Benitez {et~al.}(2002)Benitez, Maiz-Apellaniz, \&
  Canelles}]{Benitez:2002jt}
Benitez, N., Maiz-Apellaniz, J., \& Canelles, M. 2002, Phys.Rev.Lett., 88,
  081101, astro-ph/0201018

\bibitem[{Berezinskii {et~al.}(1990)Berezinskii, Bulanov, Dogiel, Ginzburg, \&
  Ptuskin}]{bookBerezinskii}
Berezinskii, V., Bulanov, S., Dogiel, V., Ginzburg, V., \& Ptuskin, V. 1990,
  Astrophysics of Cosmic Rays (North Holland)

\bibitem[{Blasi \& Amato(2012{\natexlab{a}})}]{Blasi:2011fi}
Blasi, P., \& Amato, E. 2012{\natexlab{a}}, JCAP, 1201, 010, 1105.4521

\bibitem[{Blasi \& Amato(2012{\natexlab{b}})}]{Blasi:2011fm}
------. 2012{\natexlab{b}}, JCAP, 1201, 011, 1105.4529

\bibitem[{Casse {et~al.}(2002)Casse, Lemoine, \& Pelletier}]{Casse:2001be}
Casse, F., Lemoine, M., \& Pelletier, G. 2002, Phys.Rev., D65, 023002,
  astro-ph/0109223

\bibitem[{Di~Sciascio \& Iuppa(2014)}]{DiSciascio:2014jwa}
Di~Sciascio, G., \& Iuppa, R. 2014, 1407.2144


\bibitem[{Evoli {et~al.}(2008)Evoli, Gaggero, Grasso, \&
  Maccione}]{Evoli:2008dv}
Evoli, C., Gaggero, D., Grasso, D., \& Maccione, L. 2008, JCAP, 0810, 018,
  0807.4730

\bibitem[{Fry {et~al.}(2015)Fry, Fields, \& Ellis}]{Fry:2014yqa}
Fry, B.~J., Fields, B.~D., \& Ellis, J.~R. 2015, Astrophys.J., 800, 71,
  1405.4310

\bibitem[{{Giacalone} \& {Jokipii}(1999)}]{1999ApJ...520..204G}
{Giacalone}, J., \& {Jokipii}, J.~R. 1999, \apj, 520, 204

\bibitem[{Giacinti {et~al.}(2012{\natexlab{a}})Giacinti, Kachelrie{\ss}, \&
  Semikoz}]{Giacinti:2012ar}
Giacinti, G., Kachelrie{\ss}, M., \& Semikoz, D. 2012{\natexlab{a}},
  Phys.Rev.Lett., 108, 261101, 1204.1271

\bibitem[{Giacinti {et~al.}(2013)Giacinti, Kachelrie{\ss}, \&
  Semikoz}]{Giacinti:2013wbp}
------. 2013, Phys.Rev., D88, 023010, 1306.3209

\bibitem[{Giacinti {et~al.}(2014)Giacinti, Kachelrie{\ss}, \&
  Semikoz}]{escape1}
------. 2014, Phys.Rev., D90, 041302, 1403.3380

\bibitem[{Giacinti {et~al.}(2015)Giacinti, Kachelrie{\ss}, \&
  Semikoz}]{escape2}
------. 2015, Phys.Rev., D91, 083009, 1502.01608

\bibitem[{Giacinti {et~al.}(2011)Giacinti, Kachelrie{\ss}, Semikoz, \&
  Sigl}]{Giacinti:2011uj}
Giacinti, G., Kachelrie{\ss}, M., Semikoz, D., \& Sigl, G. 2011,
  Astropart.Phys., 35, 192, 1104.1141

\bibitem[{Giacinti {et~al.}(2012{\natexlab{b}})Giacinti, Kachelrie{\ss},
  Semikoz, \& Sigl}]{Giacinti:2011ww}
------. 2012{\natexlab{b}}, JCAP, 1207, 031, 1112.5599

\bibitem[{Kachelrie{\ss} {et~al.}(2015)Kachelrie{\ss}, Neronov, \&
  Semikoz}]{LS}
Kachelrie{\ss}, M., Neronov, A., \& Semikoz, D. 2015, 1504.06472

\bibitem[{Karakula {et~al.}(1972)Karakula, Osborne, Roberts, \&
  Tkaczyk}]{Karakula:1972na}
Karakula, S., Osborne, J., Roberts, E., \& Tkaczyk, W. 1972, J.Phys., A5, 904

\bibitem[{Knie {et~al.}(1999)Knie, Korschinek, Faestermann, Wallner, Scholten,
  {et~al.}}]{Knie:1999zz}
Knie, K., Korschinek, G., Faestermann, T., Wallner, C., Scholten, J., {et~al.}
  1999, Phys.Rev.Lett., 83, 18

\bibitem[{{Lee}(1979)}]{1979ApJ...229..424L}
{Lee}, M.~A. 1979, \apj, 229, 424

\bibitem[{{Putze} {et~al.}(2011){Putze}, {Maurin}, \&
  {Donato}}]{2011A&A...526A.101P}
{Putze}, A., {Maurin}, D., \& {Donato}, F. 2011, \aap, 526, A101, 1011.0989

\bibitem[{{Shen} \& {Mao}(1971)}]{1971ApL.....9..169S}
{Shen}, C.~S., \& {Mao}, C.~Y. 1971, \aplett, 9, 169

\bibitem[{Strong \& Moskalenko(1998)}]{galprop}
Strong, A., \& Moskalenko, I. 1998, Astrophys.J., 509, 212, astro-ph/9807150

\bibitem[{Strong {et~al.}(2007)Strong, Moskalenko, \& Ptuskin}]{Strong:2007nh}
Strong, A.~W., Moskalenko, I.~V., \& Ptuskin, V.~S. 2007,
  Ann.Rev.Nucl.Part.Sci., 57, 285, astro-ph/0701517

\bibitem[{Yoon {et~al.}(2011)Yoon, Ahn, Allison, Bagliesi, Beatty,
  {et~al.}}]{Yoon:2011aa}
Yoon, Y., Ahn, H., Allison, P., Bagliesi, M., Beatty, J., {et~al.} 2011,
  Astrophys.J., 728, 122, 1102.2575

\end{thebibliography}

\end{document}